\documentclass[12pt]{amsart}
\usepackage[mathscr]{eucal}
\usepackage{amssymb,latexsym}
\usepackage{verbatim}
\usepackage{amsmath}
\usepackage{amsthm}
\usepackage{enumerate}
\usepackage{framed}

\newtheorem{thm}{Theorem}[section]

\newtheorem{Def}[thm]{Definition}
\newtheorem{rem}[thm]{Remark}
\newtheorem{prop}[thm]{Proposition}
\newtheorem{cor}[thm]{Corollary}

\newcommand\calH{{\mathcal{H}}}

\newcommand\bbP{{\mathbb P}}
\newcommand\bbR{{\mathbb R}}
\newcommand\bbZ{{\mathbb{Z}}}

\renewcommand\S{\Sigma}
\newcommand\s{\sigma}
\renewcommand\d{\partial}

\renewcommand\div{{\rm div}}

\renewcommand\th{\theta}

\DeclareFontFamily{OT1}{rsfs}{}
\DeclareFontShape{OT1}{rsfs}{m}{n}{ <-7> rsfs5 <7-10> rsfs7 <10->
rsfs10}{} \DeclareMathAlphabet{\mycal}{OT1}{rsfs}{m}{n}
\def\scri{{\mycal I}}

\newcommand\beq{\begin{equation}}
\newcommand\eeq{\end{equation}}
\newcommand\ben{\begin{enumerate}}
\newcommand\een{\end{enumerate}}
\newcommand\bit{\begin{itemize}}
\newcommand\eit{\end{itemize}}

\newcounter{mnotecount}[section]

\title
{Topological censorship \\ from the initial data point of view}

\author{Michael Eichmair, Gregory J. Galloway, and Daniel Pollack} 

\address{ETH Z\"urich \\ Departement Mathematik \\ 8092 Z\"urich, Switzerland \\}
\email{michael.eichmair@math.ethz.ch}
\thanks{Michael Eichmair gratefully acknowledges the support of NSF grant DMS-0906038 and of SNF grant 200021-140467. Gregory J. Galloway gratefully acknowledges the support of NSF grant DMS-0708048.} 

\address {University of Miami \\  Department of Mathematics \\ Coral Gables, FL 33124, U.S.A. \\}
\email {galloway@math.miami.edu}

\address {University of Washington \\ Department of Mathematics \\ Seattle, WA 98195, U.S.A. \\} 
\email {pollack@math.washington.edu}

\begin{document}
\date{}
\maketitle
\vspace{.2in}

\begin{abstract} 

We introduce a natural generalization of marginally outer trapped surfaces, called immersed marginally outer trapped surfaces, and prove that three dimensional asymptotically flat initial data sets either contain such surfaces or are diffeomorphic to $\bbR^3$. We establish a generalization of the Penrose singularity theorem which shows that the presence of an immersed marginally outer trapped surface generically implies the null geodesic incompleteness of any spacetime that  satisfies the null energy condition and which admits a non-compact Cauchy surface. Taken together, these results can be viewed as an initial data version of the Gannon-Lee singularity theorem. The first result is a non-time-symmetric version of a theorem of Meeks-Simon-Yau which implies that every asymptotically flat Riemannian $3$-manifold that is not diffeomorphic to $\bbR^3$ contains an embedded  stable minimal surface. We also obtain an initial data version of the spacetime principle of topological censorship. Under physically natural assumptions, a $3$-dimensional asymptotically flat initial data set with marginally outer trapped boundary and no immersed marginally outer trapped surfaces in its interior is diffeomorphic to $\mathbb{R}^3$ minus a finite number of open balls. An extension to higher dimensions is also discussed. 
\end{abstract}

\section{Introduction}

Topological censorship is a basic principle of spacetime physics. It is a set of results,
beginning with the topological censorship theorem of Friedman, Schleich, and 
Witt~\cite{FSW},
that establishes the topological simplicity
on
the fundamental group level of the domain of outer communications (the region outside all black holes and white holes of a spacetime) in a variety of physically natural circumstances, see e.g. \cite{FSW, Gdoc, GSWW, CGS}. Topological censorship has played an important role in black hole uniqueness theorems. In particular, it has been used to determine the topology of  black holes in $3+1$ dimensions, see e.g. \cite{ChruCosta, GSWW}. An important precursor to the principle of topological censorship is the Gannon-Lee singularity theorem \cite{Gannon, Lee}.  
All these results involve conditions that are global in time, such as global hyperbolicity or  the existence of a regular past and future null infinity. As such, they are {\it spacetime} results. It is a natural and inherently difficult question to determine whether a given initial data set gives rise to a spacetime that satisfies these conditions. 

In this paper we separate out the issue of global evolution and obtain a pure initial data version of topological censorship. 

A brief review of  topological censorship, including the result of Gannon and Lee, will be given in Section \ref{sec:intro}. In Section \ref{sec:initialdatasingularitytheorem} we take up the question of what exactly constitutes an initial data singularity theorem. We derive a generalization of the Penrose singularity theorem. The discussion leads to the notion of {\it immersed} marginally outer trapped surfaces. In Section \ref{sec:GannonLee} we present our initial data version of the Gannon-Lee singularity theorem. This result may be viewed as a non-time-symmetric version of results of  Meeks-Simon-Yau \cite{MSY}. The proof relies on recent existence results for marginally outer trapped surfaces; see \cite{AEM} and references therein.
In Section \ref{sec:initialversion} we formulate and prove an initial data version of topological censorship for $3$-dimensional initial data sets. The final section establishes a related initial data result in higher dimensions. Results of a similar flavor and related to topological obstructions to finding entire solutions of Jang's equation have been considered in \cite{SW}.

\bigskip

\noindent {\bf Acknowledgements: } We are grateful to Lars Andersson, Robert Beig, Piotr 
Chru\'sciel, Justin Corvino, and Marc Mars for their valuable comments on an earlier version of this manuscript. We also thank the referees for their careful reading of this paper and their useful suggestions. 

\section{A brief review of topological censorship} \label{sec:intro}

We expect that nontrival topological structures such as throats joining different universes pinch off and form singularities.  This intuitive idea has been captured in the following theorem proved independently by Gannon \cite{Gannon} and Lee \cite{Lee}:

\begin{thm}[\cite{Gannon, Lee}] \label{thm:GL}
Let $(M,g)$ be a globally hyperbolic spacetime which satisfies the null energy condition, ${\rm Ric}(X,X) \ge 0$ for all null vectors $X$, and which contains a Cauchy 
surface $V$ that is regular near infinity.  If $V$ is not simply connected, then $(M, g)$ is future null geodesically incomplete.
\end{thm}

Regularity near infinity is a mild asymptotic flatness condition.   Thus, under suitable curvature and causality conditions, nontrivial fundamental group entails the formation of singularities, as indicated by the future null geodesic incompleteness.

The notion of topological censorship may then be described as follows.   As the Gannon-Lee theorem suggests, nontrivial topology tends to induce gravitational collapse.  According to the weak cosmic censorship conjecture, the process of gravitational collapse leads to the formation of an event horizon which shields the singularities from view.  As a result, we expect that nontrivial topology is hidden behind the event horizon. The domain of outer communications should have simple topology. 

This notion was formalized by the topological censorship theorem of  Friedman, Schleich, and Witt \cite{FSW}.  Their theorem applies to asymptotically flat  spacetimes, i.e. spacetimes admitting a regular 
null infinity (conformal completion) $\scri = \scri^+ \cup \scri^-$, $ \scri^{\pm} \approx  \bbR \times \mathbb{S}^2$, and such that $\scri$ admits a simply connected neighborhood $U$.

\begin{thm}[\cite{FSW}]\label{topcen}
Let $(M,g)$ be a globally hyperbolic asymptotically flat spacetime that satisfies the null energy condition and such that $\scri$ admits a simply connected neighborhood $U$. Then every causal curve from $\scri^-$ to $\scri^+$ can be deformed with endpoints fixed to a curve lying in $U$.
\end{thm}

In physical terms the conclusion asserts that observers traveling from $\scri^-$ to $\scri^+$ are unable to probe any nontrivial topology.  

The domain of outer communications is the region $D  = I^-(\scri^+) \cap I^+(\scri^-)$.  The topological censorship theorem of Friedman, Schleich and Witt is really a statement about the domain of outer communications, since any causal curve from 
$\scri^-$ to $\scri^+$ is contained in $D$.  Strictly speaking, their theorem does not give any 
direct information about the topology of the domain of outer communications, because it is a statement about 
causal curves, rather than arbitrary curves.  However, in \cite{CW}, 
Chru\'sciel and Wald used the Friedman-Schleich-Witt result to prove that 
for stationary (i.e. steady state) black hole spacetimes, the domain of outer communications is simply connected, see also \cite{JV}.  Subsequent to the work of Chru\'sciel and  Wald \cite{CW}, the second-named author was able to show that the simple connectivity of the domain of outer communications holds in general:

\begin{thm}[\cite{Gdoc}] \label{thm:Gdoc} Let $(M,g)$ be an asymptotically flat spacetime such that
a neighborhood of $\scri = \scri^+ \cup \scri^-$ is simply connected.
Suppose that the domain of outer communications $D = I^-(\scri^+) \cap I^+(\scri^-)$ is globally hyperbolic and 
satisfies the null energy condition. Then $D$ is simply connected.
\end{thm}

While the proof of Theorem \ref{thm:Gdoc} makes use of the Friedman-Schleich-Witt result, the conclusion is actually stronger. Thus, in the asymptotically flat setting, topological
censorship can be taken as the statement that the domain of outer communications is simply connected.  Topological censorship has been extended in various directions, for example to the setting of asymptotically locally anti de-Sitter spacetimes \cite{GSWW},  and more recently to Kaluza-Klein spacetimes~\cite{CGS}.

\section{Initial data singularity theorems} \label{sec:initialdatasingularitytheorem}

What is an initial data singularity theorem? In view of the Penrose singularity theorem, any result that proves the existence of trapped surfaces in initial data would qualify. 
  
We recall some basic definitions.  Let $(M,g)$ be a $4$-dimensional  time-oriented Lorentzian manifold, henceforth referred to as a spacetime.  Let $(V, h, K)$ be an initial data set for $(M,g)$, that is, a spacelike hypersurface $V$ of $M$ with induced Euclidean metric $h$ and second fundamental 
form $K$.
Let $u$ denote the future pointing timelike normal to $V$ in $M$.
Let $\S$ be a closed embedded two-sided surface  in $V$.   Then $\S$ admits a smooth unit normal field
$\nu$ in $V$ that is unique up to sign. We will refer to our choice of $\nu$ as the outward pointing unit normal. 
Then $\ell_+ = u+\nu$ and $\ell_- =  u - \nu$ are, respectively, the future directed outward and inward null normal vector fields of $\S$ as a submanifold of $M$. Tracing the null second fundamental forms $\chi_{\pm}$ associated to the null normals $\ell_{\pm}$ one obtains the {\it null expansion scalars} (or {\it null mean curvatures}) $\th_{\pm} ={\rm tr} \, \chi_{\pm} = \div_{\S} \, \ell_{\pm}$.
Note that $\th_{\pm} = {\rm tr}_{\S} K \pm H$, where $H$ is the mean curvature scalar of $\S$ in $V$ with respect to $\nu$. Our sign convention for $H$ here is such that $H$ is the tangential divergence of $\nu$ in $\S \subset V$. 

In a region of spacetime where the gravitational field is strong, it can happen that $\th_- < 0$ and $\th_+ < 0$ along $\S$. If this is the case, $\S$ is called a \emph{trapped surface}. Focusing attention on the outward null normal only, $\S$ is said to be an \emph{outer trapped surface} if $\th_+ < 0$, and is said to be a \emph{marginally outer trapped surface} (MOTS) if $\th_+$ vanishes identically. 

The significance of trapped surfaces stems from their prominent role in Penrose's celebrated singularity theorem.

\begin{thm}[Penrose singularity theorem, cf.\ {\cite[Section 8.2]{HE}}]
\label{Penrose} Let $(M,g)$ be a spacetime. Assume that 
\begin{enumerate} [(i)]
\item $M$ admits a non-compact Cauchy surface $V$;
\item $M$ obeys the null energy condition;  
\item $V$ contains a trapped surface $\S$. 
\end{enumerate}
Then at least one of the future directed null normal geodesics emanating from $\S$ is incomplete.
 \end{thm}
 
Beig and \'O Murchadha \cite{Beig}  have given criteria for vacuum initial data sets to contain a trapped surface. In view  of Theorem \ref{Penrose}, we regard their result as an initial data singularity theorem.

Schoen and Yau \cite{SYbh} have given geometric conditions that imply the existence of a MOTS in an initial data set. Roughly, if enough matter is packed into a small enough 
region, then a MOTS must be present. Should conditions on an initial data set that imply the existence of a MOTS be regarded as an initial data singularity result? In view of our next result, a variant of the Penrose singularity theorem that applies to marginally outer trapped surfaces rather than trapped surfaces, the answer is yes. Related versions of the Penrose singularity theorem that apply to separating outer trapped surfaces have been considered in \cite{Gannon2, Totschnig, AMMS}. Our proof is a variation of Penrose's original argument. 
  
\begin{thm} \label{sing}
Let $(M,g)$ be a spacetime. Suppose that the following conditions are satisfied:
\begin{enumerate} [(i)]
\item $M$ admits a non-compact Cauchy surface $V$.
\item $M$ obeys the null energy condition.
\item $V$ contains a MOTS $\S$.
\item The generic condition holds on each future and past inextendible null geodesic $\eta$ normal to~$\S$.\footnote{The generic condition asks that there be a point $p$ on each null geodesic $\eta$ as in the statement of the theorem and a vector $X$ at $p$ orthogonal to $\eta'$ such that
$R(X,\eta', \eta', X)  \ne 0$. Put differently, we ask that there be a non-zero tidal acceleration somewhere along $\eta$. See Theorem 2 in Section 8.2 of \cite {HE} and Lemma 2.9 of \cite{BE}. 
}
\end{enumerate}
Then at least one of the null geodesics normal to $\S$ is future or past incomplete.

\proof  We may assume that $\S$ is connected. 
 
Assume that $\S$ separates $V$. Let $V\setminus \S  = U \cup W$ where $U, W \subset V$ are disjoint and open and where $U$ is unbounded.
By taking the time dual of $M$ if necessary, we may assume that  $\S$ satisfies  $\th_+ = 0$ with respect to the null normal  $\ell_+ = u + \nu$ where $\nu$ is the unit normal field of $\S$ in $V$ that points towards $U$. 

Consider the achronal boundary  $\partial J^+(\overline W)$.  Then
$\mathcal{H} := \partial J^+(\overline W) \setminus W$ is a $C^0$ achronal hypersurface with boundary $\S$. It is generated  by null geodesics orthogonal to $\S$ with initial tangents given by $\ell_+$. (We refer the reader to \cite{Penrose} for a thorough treatment of achronal boundaries.) 
A standard argument using the integral curves of a global timelike vector field to construct a continuous injective map from $\mathcal H$ into $\overline U$ shows that $\mathcal H$ is non-compact.  

One may now argue as in \cite[Theorem 5.2]{Fewster} 
to obtain an  affinely parametrized future inextendible null geodesic generator 
$\eta : [0, a) \to M$, $a \in (0,\infty]$,  of $\mathcal H$ such that $\eta(0) = p \in \S$ and 
$\eta'(0) = \ell_+(p)$.  Since $\mathcal H$ is achronal, there are no focal cut points to $\S$ along 
$\eta$. It follows that $\eta$  is contained in a  smooth null hypersurface $N \subset \mathcal H$  generated by null geodesics emanating from $\S$ near $p$ with initial tangents given by 
$\ell_+$; cf. Section 6 in \cite{Kemp}.

Suppose that $\eta$ is future complete, i.e., that $a = \infty$.  For each $t \ge 0$, consider
the null Weingarten map $b = b(t)$ of $N$ at $\eta(t)$ with respect to $\eta'(t)$;  
cf. e.g. \cite{Gnullmp}.
Then $b = b(t)$ satisfies the Riccati-type equation
\beq\label{riccati}
b' +b^2 + {\mathcal R} = 0,
\eeq
where a dash denotes covariant differentiation, and where $\mathcal{R}$ encodes the curvature values $R(X,\eta', \eta',Y)$ for $X,Y$ orthogonal to $\eta'$. Tracing (\ref{riccati}) we obtain the Raychaudhuri equation 
\beq\label{eq:ray}
\frac{d \th}{d t}  = - {\rm Ric}\,(\eta',\eta') - \sigma^2 - \frac1{2} \th^2   \,,
\eeq
where $\th = {\rm trace}\, b$ is the null mean curvature of $N$ along $\eta$, and where $\sigma$ is the trace of the square of the trace-free part of $b$.  

Since $\S$ is a MOTS, we have that $\th(0) = 0$.  The null energy condition shows that 
$\frac{d \th}{d t} \le 0$ for all $t \ge 0$. Standard analysis shows that 
$\th (t) \to -\infty$ in finite affine parameter time,  unless $\th = \s = 0$ along $\eta$. In the latter case, $b$ vanishes along $\eta$, so that $\mathcal R$ also vanishes by  \eqref{riccati}. This violates the generic condition. Thus 
$\eta$ is future null geodesically incomplete.

Assume now that $\S$ does not separate $V$. Then $V \setminus \Sigma$ is connected. 

In view of Theorem 1.2 in \cite{BernalSanchez}, the spacetime $(M, g)$ is isometric to $\mathbb{R} \times V$ with Lorentzian metric
$- \phi^2 dt^2 + h_t$ where $\phi$ is a smooth positive function on $ \mathbb{R} \times V$ and where $h_t$ is a Riemannian metric on $ \{t\} \times V$. Moreover, the Cauchy surface $V$ corresponds to the horizontal cross-section $\{0\} \times V$ under this isometry. Let $p : \tilde V \to V$ be a covering.  Consider the spacetime $(\tilde M, \tilde g)$ with $\tilde M =  \mathbb{R} \times \tilde V$ and $\tilde g = - \tilde \phi^2 d t^2 + \tilde h_t$ where $\tilde \phi (t, x) = \phi (t, p(x))$ and where $\tilde h_t = p^* h_t$. Clearly, $\tilde V \cong \{0\} \times \tilde V$ is a Cauchy surface for this spacetime and $P : \tilde M \to M$ given by $(t, x) \mapsto (t, p(x))$ is a covering map.

Since $\Sigma$ is two-sided but does not separate $V$ we can make a cut along $\Sigma$ to obtain a connected manifold $V'$ with two boundary components, each isometric to $\Sigma$. Taking $\mathbb Z$ copies of $V'$ and gluing these copies end-to-end we obtain a covering $p: \tilde V \to V$ of $V$. The inverse image $p^{-1}(\Sigma)$ consists of $\mathbb Z$ copies of $\Sigma$, each one separating $\tilde V$.  Let $\Sigma_0$ denote one
of these copies.   As per the comment above, we obtain a covering spacetime $\tilde M$
with non-compact Cauchy surface  $\tilde V$. The curvature assumptions on $M$ lift to $\tilde M$. It follows from our earlier arguments that there exists a null geodesic $\eta_0$ normal to $\S_0$ that is future or past incomplete.  The projection of $\eta_0$ under
the covering map $P$ is a past or future incomplete null geodesic normal to $\S$ 
in $M$.~\qed
 \end{thm}

\begin{rem}
{\rm
Theorem \ref{sing} is false without the generic condition. This can be seen in the example of the extended Schwarzschild solution where every Cauchy surface intersects the event horizon in a MOTS whose outward null normal geodesics are the generators of the future horizon. These geodesics are known to be complete. The generic condition fails along each such null geodesic. This rigidity is characteristic in the setting of Theorem \ref{sing} when the generic condition fails while future null geodesic completeness holds: One of the families of future or past null geodesics emanating from the MOTS forms a totally geodesic null hypersurface and the generic condition fails along all of the null geodesics in this family; cf. Theorem \ref{rigid} in the appendix. Thus, we may still conclude null-geodesic incompleteness of the spacetime if we weaken the hypotheses of Theorem \ref{sing} to only require that the generic condition hold along one inward and one outward future pointing null normal geodesic ray and one inward and one outward past pointing null normal geodesic ray. 
}
\end{rem}

Our view of what should be considered an initial data singularity result  must be taken one step further to accommodate other fundamental examples. There is a more general type of object in an initial data set that implies a Penrose-type singularity theorem, which we refer to as 
an {\it immersed} MOTS.  

\begin{Def}
Let $(V, h, K)$ be an initial data set. A subset $\S \subset V$ is called an \emph{immersed marginally outer trapped surface} if there exists a finite covering $p: \tilde V \to V$ of $V$ and a closed marginally outer trapped surface $\tilde \S$ in $(\tilde V, \, p^{*}h, \, p^{*}K)$ such that $p(\tilde \S) = \S$. 
\end{Def}

We say that $\S$ is a spherical (toroidal, etc.) immersed MOTS if $\tilde \S$ is spherical (toroidal, etc.).

\begin{cor}\label{sing2}
Theorem \ref{sing} remains valid when the condition that $V$ contain a MOTS is replaced by the condition that $V$ contain an immersed MOTS. 
\proof Let $p : \tilde V \to V$ be a cover of $V$ and let $\tilde \Sigma \subset \tilde V$ be a MOTS in $(\tilde V, \, p^{*}h, \, p^{*}K)$. Let $P : \tilde M \to M$ be the associated covering spacetime. This spacetime satisfies the hypothesis of Theorem \ref{sing} relative to the MOTS  $\tilde \S \subset \tilde V$. Theorem \ref{sing} then asserts the existence of an incomplete null geodesic normal to $\tilde \S$.  We then project this geodesic to $M$ using $P$. ~\qed 
\end{cor}

The best known example of an immersed MOTS (that is not a MOTS) occurs in the so-called $\bbR\bbP^3$ geon; see e.g. \cite{FSW} for a detailed description.  
The $\bbR\bbP^3$ geon is a globally hyperbolic spacetime that is double covered by the extended Schwarzschild spacetime.  Its Cauchy surfaces have the topology of a once punctured $\bbR\bbP^3$. The Cauchy surface that is covered by the $t=0$ slice in the extended Schwarzschild spacetime contains a projective plane $\S$ that is covered by the unique minimal sphere in the Schwarzschild slice. Hence $\S$ is a spherical immersed MOTS. 

An immersed MOTS does not cover an embedded submanifold in $V$ in general. To illustrate this point, we consider
the following $2+1$-dimensional toy model of the $\bbR\bbP^3$ geon.

Consider the manifold $\tilde M = \bbR^2 \times \mathbb{S}^1$ with coordinates $(t, y, \th)$ where $t, y \in \bbR$ and $\th \in \mathbb{R} / {2 \pi \mathbb{Z}}$ and with Lorentzian metric
$\tilde g = -dt^2 + dy^2 + d\th^2$. The slice $t= 0$ is a Cauchy surface for the spacetime  $(\tilde M, \tilde g)$. Geometrically, it is a flat cylinder.  Let $M$ be the manifold obtained from $\tilde M$ by identifying points via the involution $(y,\th) \to (-y,\th + \pi)$.  $M$ may be described as the subset $\{y \ge 0\}$ of $\tilde M$, with the points $(t,0, \th)$ and $(t, 0, \th + \pi)$ in the timelike surface $y = 0$   identified. Since $\tilde g$ is invariant  under the involution, it descends to a flat Lorentzian metric $g$ on $M$. Hence 
$(\tilde M, \tilde g)$ is a double covering of $(M,g)$.  The slice $t = 0$ in $M$ is a Cauchy surface for $(M,g)$. It has the topology of a punctured projective plane. We now perturb the slice $t = 0$ in $\tilde M$.  For $a > 0$ sufficiently small, the surface $$\tilde S = \{ (a \sin(2 \theta), y, \theta) : y \in \mathbb{R} \text{ and } \theta \in \mathbb{R} /  2 \pi \mathbb{Z} \} \subset \tilde M$$ is a Cauchy surface for $(\tilde M,\tilde g)$.  The intersection $\tilde\S$ of $\tilde S$  with the null hypersurface  $t = y$ is a  circular MOTS, which may be viewed as a  perturbation of the  MOTS at the intersection of $t = 0$ and $t =y$.  Since $\tilde S$ is invariant under the involution, it descends to a Cauchy surface $S$ of $(M,g)$.  Then $\S = p(\tilde \S)$ is an immersed MOTS in $S$. It has transverse self-intersections at the points $(0,0,0)$ and $(0,0, \pi/4)$.

\section{An initial data version of the Gannon-Lee singularity theorem} \label{sec:GannonLee}

Let $(V,h,K)$ be an $n$-dimensional asymptotically flat initial data set. This means that $V$ is connected and that the complement of some compact subset has finitely many components, each of which is diffeomorphic to $\bbR^n \setminus B_1(0)$, and such that  $h_{ij}-\delta_{ij}$ and $K_{ij}$ decay suitably in Euclidean coordinates. Decay conditions that ensure that large coordinate spheres are null mean convex as in \cite[(1)]{E2} are sufficient for our applications.

\begin{thm}\label{glid}
Let $(V, h, K)$ be a $3$-dimensional asymptotically flat initial data set without boundary. If $V$ is not diffeomorphic to $\bbR^3$ then $V$ contains an immersed MOTS. 
\end{thm} 

Theorem \ref{sing} and Theorem \ref{glid} together imply that a spacetime is singular if it satisfies appropriate curvature conditions and if it contains an asymptotically flat and topologically nontrivial Cauchy surface. We view Theorem \ref{glid} as an initial data version of the Gannon-Lee singularity theorem (Theorem \ref{thm:GL}).

For time-symmetric data, MOTSs are the same as minimal surfaces. In this special case, the conclusion of Theorem \ref{glid} follows from the work of Meeks-Simon-Yau \cite{MSY}, who prove that an asymptotically flat $3$-manifold that is not diffeomorphic to $\bbR^3$ contains a stable minimal sphere or projective plane. 

Beig and \'O Murchadha \cite{Beig91}, Miao \cite{miao}, and Yan \cite{Yan:2005}  have constructed examples of asymptotically flat scalar flat metrics on $\bbR^3$ that admit stable minimal spheres. In particular, initial data may well contain MOTSs and be diffeomorphic to $\bbR^3$ at the same time. 

In the proof of Theorem \ref{glid} we rely on the positive resolution of the Poincar\'e and Geometrization conjectures, see \cite{Perel1, Perel3, Perel2} and \cite{3BMP, CZ, KL, MT1, MT2}, and on the work of Hempel \cite{Hempel} on the structure of fundamental groups of closed three-manifolds. We are grateful to Ian Agol for valuable correspondence on the underlying geometric group theory. 

The following existence result for closed MOTSs was proposed by R. Schoen and is based on forcing a blow up of Jang's equation and analysis of the blow up set as in \cite{SY2}. It was first proved by Andersson and Metzger \cite{AM2} in the $3$-dimensional case and then by the first-named author \cite{E1, E2} in general dimension (with small singular set in dimension $n \geq 8)$ using different methods to force and control the blow up. The survey article \cite{AEM} contains an extensive overview of the techniques developed in \cite{SY2, AM2, E1, E2}, including a discussion of the geometric properties of the MOTS whose existence is established. 

\begin{thm}[Cf. Theorem 3.3 in \cite{AEM}]\label{exist}  Let $(V, h, K)$ be an $n$-dimensional initial data set where $3 \leq n \leq 7$. Let $W \subset V$ be a connected compact $n$-dimensional submanifold with boundary. Assume that its boundary $\d W$ can be expressed as a disjoint union of closed hypersurfaces $\d W = \S_{in} \cup \S_{out}$ such that $\th^+ < 0$ along $\S_{in}$ with respect to the null normal whose projection  points into $W$ and such that $\th^+ > 0$  along $\S_{out}$ with respect to the null normal whose projection  points out of $W$.  Then there exists a closed MOTS $\S$ in $W$ that separates $\S_{in}$ from $\S_{out}$. This MOTS is almost minimizing.
\end{thm}
 
\begin{rem} \label{rem:perturbation} {\rm Assume that the initial data set in Theorem \ref{exist} satisfies the dominant energy condition. This means that $\mu \ge |J|$ on $V$ where $\mu = \frac{1}{2} (R - |K|^2 + (\text{trace} (K))^2)$ and $J = \div (K - \text{trace} (K) h)$. Then there exists a MOTS $\Sigma \subset W$ as in the conclusion of the theorem that admits a metric of positive scalar curvature. In particular, when $n = 3$, there exists a spherical MOTS in $W$. To see this, note that the induced metric on any closed stable MOTS is conformal to a metric of non-negative constant scalar curvature by Theorem 2.1 in \cite{GS}. If the dominant energy condition $\mu \geq |J|$ is strict at any point on $\Sigma$, then the induced metric is conformal to a metric of constant positive scalar curvature. We can apply Theorem 18 in \cite{eichetal} to find a sequence of initial data $(h_i , K_i)$ on $V$ converging to the original data $(h, K)$ as $i \to \infty$ and such that $(V, h_i, K_i)$ satisfies the strict dominant energy condition $\mu_i > |J_i|$. It follows that the MOTS $\Sigma_i \subset W$ in $(V, h_i, K_i)$ in the conclusion of Theorem \ref{exist} admits a metric of positive scalar curvature. The uniform almost minimizing property of $\Sigma_i$ shows that a subsequence of $\Sigma_i$ converges smoothly to a closed MOTS $\Sigma \subset W$ in $(V, h, K)$ that separates $\S_{in}$ from $\S_{out}$.}
\end{rem}
 
Let $(V,h,K)$ and $W$ be as in the statement of Theorem \ref{exist}.  We say that the boundary $\d W$ is (strongly) {\it null mean convex} if it has positive outward null expansion, $\th^+ > 0$, and negative inward null expansion, $\th^-< 0$.  Round spheres in Euclidean slices of Minkowski space and, more generally, large radial spheres in asymptotically flat initial data sets are null mean convex.

\begin{prop}\label{connected} 
Let $(V,h,K)$  be an $n$-dimensional initial data set where $3 \leq n \leq 7$. Let $W \subset V$ be a connected compact $n$-dimensional submanifold with null mean convex boundary. The boundary of $W$ is connected if there are no MOTSs in $W$.
\proof Suppose $\d W$ is not connected.  We designate one component of $\d W$ as  $\S_{in}$ and the union of the others as $\S_{out}$.  Then $\S_{in}$ and  $\S_{out}$ satisfy the null expansion conditions of Theorem \ref{exist}. In particular, there exists a MOTS $\S$ in $W$. ~\qed
\end{prop}

\proof[Proof of Theorem \ref{glid}]  
Assume that there are no immersed MOTSs in $V$. By suitably truncating the ends of $V$ we obtain a compact connected $3$-dimensional submanifold $W \subset V$ with null mean convex boundary $\d W$ whose components correspond to the ends of $V$. If $V$ has more than one end, then $W$ contain a closed MOTS by Proposition \ref{connected}. This contradicts our assumption that $(V, h, K)$ contains no immersed MOTSs. Thus $V$ has exactly one end. 
  
If $V$ is not orientable, we can pass to the orientable double cover $p: \tilde V \to V$. Then $(\tilde V, p^* h, p^* K)$ is an asymptotically flat initial data set with two ends and thus contains a MOTS. This contradicts our assumption that $(V, h, K)$ contains no immersed MOTSs. Thus $V$ is orientable. 

It follows that we can express $V$ as a connected sum $V = \bbR^3 \# N$ where $N$ is a compact orientable $3$-manifold. Note that $\pi_1(V) = \pi_1(N)$. 

The work of Hempel \cite{Hempel} in conjunction with the positive resolution of the geometrization conjecture shows that $\pi_1(N)$ is a residually finite group, i.e. for every non-identity element in the group there is a normal subgroup with finite index that does not contain that element.
In particular, if $V$ is not simply connected, then $\pi_1(V)$ contains a proper normal subgroup with finite index $k >1$. Such a subgroup gives rise to a $k$-sheeted covering $p:  \tilde V \to V$ of $V$. Note that $(\tilde V,\,  p^* h, \, p^*K)$ is an asymptotically flat initial data set with $k$ ends. In particular, it contains a MOTS. This contradicts our assumption that $(V, h, K)$ contains no immersed MOTS. Thus $V$ is simply connected. 

By the positive resolution of the Poincar\'e conjecture, $N$ is diffeomorphic to $\mathbb{S}^3$ and $V$ is diffeomorphic to $\mathbb{R}^3$. ~\qed

\begin{rem} \label{rem:anotherone} {\rm Assume that $(V, h, K)$ satisfies the dominant energy condition. In view of Remark \ref{rem:perturbation}, the conclusion of Proposition \ref{connected} can be sharpened as follows. If there is no spherical MOTS in $W$, then $\d W$ is connected. This leads to a stronger conclusion in Theorem \ref{glid}. Either $V$ is diffeomorphic to $\mathbb{R}^3$ or $(V, h, K)$ contains a spherical immersed MOTS.} 
\end{rem}


\section{An initial data version of topological censorship} \label{sec:initialversion}

As in  Theorem \ref{topcen}, consider the domain of outer communications 
$D  = I^-(\scri^+) \cap I^+(\scri^-)$ in a regular black hole spacetime $(M, g)$ that satisfies the null energy condition. Assume that $D$ is globally hyperbolic and consider a Cauchy surface for $D$ whose closure $V$ intersects the event horizon in a compact surface.  It is shown in \cite{HE, Wald} that there can be no trapped surface in $V \setminus \d V$ as otherwise it would be visible at $\scri^+$, which is not possible. This remains true for marginally trapped surfaces, i.e. surfaces for which $\th_+$ and $\th_-$ are nonpositive;
cf. \cite[Proposition 12.2.2]{Wald} and \cite[Theorem 6.1]{CGS}. An argument similar to that of the proof of Theorem \ref{sing} shows that there can be no immersed MOTS in $V\setminus \d V$ either; cf. \cite[Proposition 12.2.4]{Wald} and \cite[Remark 6.5]{CGS}.

For an initial data version of topological censorship, we think of the initial data $(V, h, K)$ as representing an asymptotically flat slice in the domain of outer communications. The boundary $\d V$ is thought of as a cross section of the event horizon. This is reflected in the requirement that $\d V$ be a MOTS, possibly with multiple components.

\begin{thm}\label{topcenid} 
Let $(V, h, K)$ be a $3$-dimensional asymptotically flat initial data set. Assume that every component of $\partial V$ is a MOTS, either with respect to the unit normal pointing into $V$ or the unit normal pointing out of $V$. If there are no immersed MOTSs in $V \setminus \partial V$, then $V$ is diffeomorphic to $\bbR^3$ minus a finite number of open balls.

\proof  
The proof is similar to that of Theorem \ref{glid}. 

By suitably truncating the asymptotically flat ends of $V$ we obtain a compact connected $3$-manifold $W \subset V$ whose boundary is the disjoint union of $\partial V$ and a null mean convex surface whose components correspond to the asymptotically flat ends of $V$.  

Assume that $V$ has more than one end. We may split $\partial W$ into non-empty unions of components $\S_{out}$ and $\S_{in}$ such that $\th_+ \leq 0$ along $\S_{in}$ with respect to the unit normal pointing into $W$ and with strict inequality on some component, and such that $\theta_+ \geq 0$ along $\S_{out}$ with respect to the unit normal pointing out of $W$ and strict inequality on some component. Theorem \ref{exist} still applies in this situation,  cf. Section 5 in \cite{AM2} or Remark 4.1 in \cite{E2}, and shows that there exists a MOTS $\Sigma \subset W$ that separates $\S_{out}$ from $\S_{in}$. At least one component of $\Sigma$ must be disjoint from $\partial V$ and is thus contained in $V \setminus \d V$. This contradicts our assumptions. It follows that $V$ has only one end. 

The same covering argument as in the proof of Theorem \ref{glid} shows that $V$ is orientable.  Moreover, if some component of $\d V$ is not a $2$-sphere, then $V$ admits a double covering (cf. \cite[Lemma 4.9]{Hempel1}). Such a covering has two ends and thus contains a MOTS in its interior, contradicting our assumption. It follows that the components of $\d V$ are spherical.

We see that $V$ can be expressed as a connected sum  $V = \bbR^3 \# N$ where $N$ is a compact orientable $3$-manifold with boundary $\d N$. Let $\hat N$ be the smooth closed orientable $3$-manifold obtained from $N$ by gluing in balls along each of the spherical components of $\partial N$. 

Assume that $\hat N$ is not simply connected. As in the proof of Theorem \ref{glid}, we see that $\hat N$ admits a $k$-sheeted covering for some $1 < k < \infty$. Since balls and their complements in $\bbR^3$ are simply connected, it follows that $V$ admits a $k$-sheeted covering $p : \tilde V \to V$. The boundary of $\tilde V$ covers the boundary of $V$. It is a MOTS with respect to the pull-back data $(p^*h, p^* K)$. Hence $(\tilde V, \, p^* h, \, p^* K)$ is an asymptotically flat initial data set with MOTS boundary and $k$ ends. As above, we see that $(\tilde V, \, p^* h, \, p^* K)$ contains a closed MOTS that is disjoint from $\partial \tilde V$. It follows that $V \setminus \partial V$ contains an immersed MOTS, contrary to our assumption. Thus $\hat N$ is simply connected and hence, by the positive resolution of the Poincar\'e conjecture, diffeomorphic to $\mathbb{S}^3$. 

It follows that $V = \bbR^3 \# N$ is diffeomorphic to $\bbR^3$ minus a finite number of open balls.~\qed
\end{thm}

\section{Higher dimensions}

The proofs of Theorem \ref{glid} and Theorem \ref{topcenid} rely on facts that are specific to dimension three, specifically the positive resolution of the Geometrization conjecture. In this section we present a simple topological condition that implies the existence of immersed MOTSs in asymptotically flat initial data sets of  dimension up to and including seven.
  
\begin{thm}\label{nglid} Let $(V, h, K)$ be an $n$-dimensional asymptotically flat initial data set without boundary where $3\le n \le 7$. If $V$ contains a closed non-separating hypersurface, then $V$ contains an immersed MOTS. 

\proof The argument in the proof of Theorem \ref{glid} shows that $V$ has exactly one end and that $V$ is orientable. Let $\Sigma$ be a closed non-separating hypersurface in $V$. We can double $V$ along $\Sigma$ to obtain a two sheeted covering $p : \tilde V \to V$ of $V$. Note that $(\tilde V, \, p^* h, \, p^* K)$ is oriented and has two asymptotically flat ends. The same argument as in the proof of Theorem \ref{glid} shows that $(\tilde V, \, p^* h, \, p^* K)$ contains a closed  MOTS. This MOTS projects to an immersed MOTS in $V$. \qed
\end{thm}

The existence of a closed non-separating hypersurface in Theorem \ref{nglid} follows from the non-vanishing of the first Betti number of $V$. We briefly sketch the standard topological argument. We assume that $V$ is oriented. Let $\hat V$  be the smooth compact orientable manifold obtained from a one point compactification of $V$. Then $H_1(\hat V,\bbZ) = H_1(V,\bbZ)$ so that $b_1(\hat V) > 0$.  Poincar\'e duality and the fact that there is no torsion in co-dimension one homology implies  that $b_1(\hat V) > 0$ if and only if $H_{n-1}(\hat V, \bbZ) \ne 0$. By Thom's realizability theorem, see e.g. Theorem 11.16 in \cite{Bredon}, we may choose a non-vanishing class in $H_{n-1}(\hat V, \bbZ)$ and represent it by a closed hypersurface. There is a component $\hat \S$ of this hypersurface that does not separate $\hat V$. We may assume that $\hat \S$ does not pass through the point at infinity of $\hat V$ and hence view $\hat \Sigma$ as a smooth closed hypersurface $\Sigma$ of $V$. Clearly, $\Sigma$ does not separate.

In conclusion, we arrive at the following corollary.

\begin{cor} Let $(V, h, K)$ be an $n$-dimensional asymptotically flat initial data set without boundary where $3\le n \le 7$. If $b_1(V) > 0$ then $V$ contains an immersed MOTS. 
\end{cor}

\section{Appendix} \label{sec:rigidity}

\begin{thm} \label{rigid} Let $(M,g)$ be a future null geodesically complete spacetime that satisfies the null energy condition and which contains a non-compact Cauchy surface $V$. Assume that there exists a closed, connected, separating hypersurface $\S \subset V$ that is a MOTS with respect to the future directed null normal that points towards an unbounded component of $V \setminus \S$. The future inextendible null geodesics emanating from $\S$ in the direction of this null normal form a smooth totally geodesic null hypersurface. 

\proof 
Let $V\setminus \S  = U \cup W$ where $U, W \subset V$ are disjoint and open and where $U$ is unbounded. Let $\ell_+$ be the future directed null normal that points towards $U$. Consider the $C^0$ achronal hypersurface $\calH = \partial J^+(\overline W) \setminus W$. It is generated by null geodesic segments orthogonal to $\S$ with initial tangents given by $\ell_+$. As in  the proof of Theorem \ref{sing}, $\calH$ is non-compact and contains the trace of a null geodesic
$\eta : [0, \infty) \to M$ such that $\eta(0) = p \in \S$ and  $\eta'(0) = \ell_+(p)$.
Let $S_1$ be the component of  $\d I^-(\eta) \cap I^+(V)$ containing $\eta((0,\infty))$. Standard arguments show that $S_1$ is an achronal $C^0$ hypersurface generated by future complete null geodesics, cf. \cite{Penrose} and the proof of Theorem IV.1 in \cite{Gnullmp}. Lemma IV.2 in \cite{Gnullmp} implies that $S_1$ has null mean curvature $\theta_1 \ge 0$ in the support sense; cf. \cite[Definition III.2]{Gnullmp}. In fact, sets of the form $\d I^-(x)$ where $x \in S_1$ lies on a generator through and to the future of $y \in S_1$ provide the required family of locally smooth past support null hypersurfaces at $y$. Let $S_2 = \calH \setminus (\S \cup {\rm Cut}(\S))$ where 
${\rm Cut}(\S)$ is the set of null focal cut points to $\S$ in direction $\ell_+$. Now $S_2$ is a smooth null hypersurface generated by null geodesics in direction of $\ell_+$ up to (but not including) cut points.  Equation \eqref{eq:ray} and the null energy condition imply that $S_2$ has null mean curvature $\th_2 \le 0$. 

Causal considerations show that $S_2$ lies locally to the future of $S_1$ near points where they intersect. In particular, this holds along $\eta ((0,\infty))$. The maximum principle for $C^0$ null hypersurfaces, Theorem III.4 in \cite{Gnullmp}, together with a straightforward continuation argument, implies that $S = S_1 = S_2$ and that $S$ has null mean curvature identically equal to zero. It follows from equation \eqref{eq:ray} that the Weingarten map of $S$ vanishes  identically.  \qed
\end{thm}

Theorem \ref{rigid} shows that $(M, g)$ is future null geodesically incomplete if the generic condition is violated along a single null geodesic in direction $\ell_+$. 
 
\bibliographystyle{amsplain}

\providecommand{\bysame}{\leavevmode\hbox to3em{\hrulefill}\thinspace}

\end{document}